\documentclass[fleqn,usenatbib,a4paper]{mnras}

\usepackage{savesym}
\usepackage{amsmath}
\savesymbol{iint}
\usepackage{txfonts}
\restoresymbol{TXF}{iint}

\usepackage[T1]{fontenc}
\usepackage{ae,aecompl}

\usepackage{graphicx}	
\usepackage{amssymb}	
\usepackage{xspace}
\usepackage{enumitem}
\usepackage{booktabs}

\newcommand{\chandra}{{\it Chandra}\xspace}
\newcommand{\hst}{{\it HST}\xspace}

\def\ergs{\hbox{${\rm erg\, s^{-1}}$}\xspace}
\def\kms{\hbox{${\rm km\, s^{-1}}$}\xspace}
\def\cts{\hbox{${\rm counts\, s^{-1}}$}\xspace}

\def\Msun{\hbox{$\thinspace M_{\odot}$}\xspace}
\def\lk{\hbox{$\thinspace L_{K}$}\xspace}

\def\LKsun{\hbox{$\thinspace L_{K\odot}$}\xspace}

\def\lx{\hbox{$\thinspace L_{x}$}\xspace}

\title[X-ray populations of NGC 7457]{Deep Chandra observations of NGC~7457, the X-ray point source populations of a low mass early-type galaxy}

\author[M. B. Peacock et al.]{
Mark B. Peacock$^{1}$\thanks{E-mail: mpeacock@msu.edu (MBP)} 
Stephen E. Zepf$^{1}$ 
Arunav Kundu$^{2}$ 
Thomas J. Maccarone$^{3}$
\newauthor Bret D. Lehmer$^{4}$
Anthony H. Gonzalez$^{5}$
and Claudia Maraston$^{6}$
\\
$^{1}$Department of Physics and Astronomy, Michigan State University, East Lansing, MI 48824, USA \\
$^{2}$Eureka Scientific, Inc., 2452 Delmer Street, Suite 100 Oakland, CA 94602, USA \\
$^{3}$Texas Tech University, Physics Department, Box 41051, Lubbock, TX 79409, USA \\
$^{4}$Department of Physics, University of Arkansas, 226 Physics Building, 835 West Dickson Street, Fayetteville, AR 72701, USA \\
$^{5}$Department of Astronomy, University of Florida, Gainesville, FL 32611, USA \\
$^{6}$Institute of Cosmology and Gravitation, Dennis Sciama Building, Burnaby Road, Portsmouth PO1 3FX, UK
}

\date{Accepted XXX. Received YYY; in original form ZZZ}

\pubyear{2017}

\begin{document}
\label{firstpage}
\pagerange{\pageref{firstpage}--\pageref{lastpage}}
\maketitle

\begin{abstract}
\label{sec:abstract}

We present the X-ray point source population of NGC~7457 based on 124~ks of \chandra observations. Previous deep \chandra observations of low mass X-ray binaries (LMXBs) in early-type galaxies have typically targeted the large populations of massive galaxies. NGC~7457 is a nearby, early-type galaxy with a stellar luminosity of $1.7\times10^{10}\LKsun$, allowing us to investigate the populations in a relatively low mass galaxy. We classify the detected X-ray sources into field LMXBs, globular cluster LMXBs, and background AGN based on identifying optical counterparts in new \hst/ACS images. We detect 10 field LMXBs within the $r_{ext}$ ellipse of NGC~7457 (with semi-major axis $\sim$ 9.1~kpc, ellipticity = 0.55). The corresponding number of LMXBs with $\lx>2\times10^{37}\ergs$ per stellar luminosity is consistent with that observed in more massive galaxies, $\sim 7$ per $10^{10} \LKsun$. We detect a small globular cluster population in these \hst data and show that its colour distribution is likely bimodal and that its specific frequency is similar to that of other early type galaxies. However, no X-ray emission is detected from any of these clusters. Using published data for other galaxies, we show that this non-detection is consistent with the small stellar mass of these clusters. We estimate that 0.11 (and 0.03) LMXBs are expected per $10^{6}\Msun$ in metal-rich (and metal-poor) globular clusters. This corresponds to 1100 (and 330) LMXBs per $10^{10} \LKsun$, highlighting the enhanced formation efficiency of LMXBs in globular clusters. A nuclear X-ray source is detected with \lx varying from $2.8-6.8\times10^{38}\ergs$. Combining this \lx with a published dynamical mass estimate for the central SMBH in NGC~7457, we find that $\lx/L_{Edd}$ varies from $0.5-1.3\times10^{-6}$. 

\end{abstract}




\section{Introduction}
\label{sec:intro}

The superb spatial resolution of the \chandra X-ray observatory has revealed large point source populations in local galaxies \citep[see e.g. the review of ][and references therein]{Fabbiano06}. For the old stellar populations of early-type galaxies these sources are primarily thought to be low mass X-ray binaries (LMXBs). For the closest galaxies, deep \chandra observations can probe the LMXB populations to X-ray luminosities (\lx) of $few \times10^{37} \ergs$, or fainter. This allows us to study a significant fraction of the active LMXB population and investigate the X-ray luminosity function (XLF), the shape of which may provide clues to the nature of the LMXB populations. 

Beyond the local group, optical emission from LMXBs will be too faint to detect with \hst observations. However, combining \chandra with optical \hst data is powerful for identifying globular cluster LMXBs and background active galactic nuclei (AGN), by detecting their host clusters and galaxies, respectively. Such work has demonstrated that a large fraction ($20-70\%$) of the LMXBs observed in nearby early-type galaxies are located in their globular clusters \citep[e.g.][]{Angelini01, Kundu02, Jordan04}. These globular cluster LMXBs are likely formed via dynamical interactions \citep[e.g.][]{Clark75, Verbunt87, Jordan07, Peacock09} and should therefore be treated as a distinct population from the LMXBs in the field of a galaxy \citep[although a small fraction of field LMXBs may have originated in clusters, e.g.][]{Kim09}. 

Deep \chandra observations of massive early-type galaxies have previously been used for detailed studies of their LMXB populations \citep[e.g.][]{Sivakoff08, Brassington08, Brassington09, Voss09, Li10, Luo13, Lehmer14, Lin15}. However, the available data for lower mass early-type galaxies is significantly less deep \citep[e.g.][who studied six galaxies with $\lk=1-2\times10^{10}\LKsun$, but with relatively bright detection limits of $\lx>1\times10^{38}\ergs$]{Coulter16}. In this paper we present the field LMXB population of NGC~7457 based on deep \chandra observations. This is a nearby \citep[$d=12.1~Mpc$;][]{Tully13} early-type galaxy \citep[S0;][]{deVaucouleurs91} that has a relatively low mass, with a velocity dispersion of $75\kms$ \citep{Cappellari13b} and K-band luminosity of $1.7\times10^{10}\LKsun$ \citep{Jarrett03}. It also has a central super massive black hole (SMBH) with dynamical mass estimates that suggest it is relatively low mass \citep[$M_{BH}=4.1^{+1.2}_{-1.7} \times10^{6}\Msun$;][]{Gebhardt03, Gultekin09a}, consistent with galaxy's low stellar mass.

We present these new \chandra observations of NGC~7457 and the detected X-ray populations in Section \ref{sec:xray}. In Section \ref{sec:optical}, we present \hst observations of the galaxy, which we use to identify and study globular cluster candidates and classify sources into field LMXBs, globular cluster LMXBs and background AGN. In Sections \ref{sec:lmxb} and  \ref{sec:gc_lmxb}, we discuss the observed field and globular cluster LMXB populations and compare them with those observed in other, more massive, early-type galaxies. In Section \ref{sec:agn} we discuss the observed nuclear emission, before concluding in Section \ref{sec:conclusions}.

\section{\chandra observations \& reduction} 
\label{sec:xray}

\begin{table*}
 {\centering
 \footnotesize
 \caption{Detected sources X-ray sources within $r_{ext}$ \label{tab:data} }
   \begin{tabular}{lccrrrrrrrrrc}
   \hline 
  \hline
  \noalign{\vskip 1mm}    
   &    &    &  \multicolumn{2}{c}{OBS ID 11786} & \multicolumn{2}{c}{OBS ID 17007} & \multicolumn{2}{c}{OBS ID 18440} & \multicolumn{2}{c}{COMBINED} &  &  \\
\cmidrule(r{5pt}){4-5}\cmidrule(r{5pt}){6-7}\cmidrule(r{5pt}){8-9}\cmidrule(r{5pt}){10-11}
ID & RA & DEC     &  RATE$^{a}$ & $L_{x}$$^{b}$ & RATE$^{a}$  & $L_{x}$$^{b}$ & RATE$^{a}$ &      $L_{x}$$^{b}$        & RATE$^{a}$ &       $L_{x}$$^{b}$ [+ / - err] & $R_{gcp}$$^{c}$ & $f_{opt}^{d}$ \\ 
    \noalign{\vskip 1mm}    
 \hline
    \noalign{\vskip 1mm}    
01* & 345.2496621 & 30.1448416 & 16.7 & 2.39  & 40.5 & 7.24 & 33.8 & 6.04 & 32.3 & 5.46 +0.50 -0.49 & 0.2 & Nuc. \\
02 & 345.2509550 & 30.1463913 & 1.4 & 0.20  & 1.1 & 0.20 & 0.3 & 0.06 & 0.8 & 0.14 +0.11 -0.06 & 6.8 & $<$10$\arcsec$ \\
03 & 345.2449162 & 30.1428012 & 1.5 & 0.21  & 1.5 & 0.26 & 1.2 & 0.22 & 1.4 & 0.23 +0.13 -0.08 & 16.5 & no \\
04 & 345.2549238 & 30.1500442 & 12.7 & 1.90  & 11.9 & 2.32 & 16.8 & 3.15 & 14.1 & 2.53 +0.34 -0.33 & 24.8 & no \\
05 & 345.2419307 & 30.1472114 & 6.5 & 0.93  & 6.0 & 1.07 & 6.8 & 1.21 & 6.4 & 1.09 +0.24 -0.19 & 25.4 & no \\
06 & 345.2464571 & 30.1519551 & 6.9 & 0.99  & 3.2 & 0.57 & 2.8 & 0.50 & 3.9 & 0.66 +0.19 -0.14 & 27.3 & no \\
07 & 345.2613643 & 30.1405110 & 0.7 & 0.10  & 1.4 & 0.28 & 1.5 & 0.26 & 1.3 & 0.22 +0.13 -0.08 & 39.8 & no \\
08 & 345.2654915 & 30.1451776 & 3.4 & 0.49  & 0.0 & 0.00 & 0.0 & 0.00 & 1.2 & 0.20 +0.08 -0.05 & 49.4 & no \\
09 & 345.2658659 & 30.1370842 & 1.4 & 0.21  & 0.9 & 0.17 & 0.6 & 0.10 & 0.9 & 0.15 +0.11 -0.06 & 57.8 & no \\
10 & 345.2550423 & 30.1656542 & 2.2 & 0.32  & 0.2 & 0.03 & 0.4 & 0.07 & 0.7 & 0.13 +0.10 -0.05 & 76.6 & yes \\
11 & 345.2666487 & 30.1249590 & 0.3 & 0.04  & 0.4 & 0.07 & 1.4 & 0.26 & 0.8 & 0.14 +0.10 -0.06 & 89.2 & no \\
12 & 345.2199592 & 30.1481165 & 1.5 & 0.21  & 3.2 & 0.58 & 2.8 & 0.51 & 2.6 & 0.45 +0.17 -0.12 & 93.1 & yes \\
13 & 345.2698244 & 30.1239772 & 5.7 & 0.83  & 1.1 & 0.21 & 2.1 & 0.38 & 2.6 & 0.45 +0.16 -0.11 & 98.1 & yes \\
14 & 345.2170819 & 30.1511690 & 1.8 & 0.27  & 0.4 & 0.13 & 1.9 & 0.42 & 1.4 & 0.30 +0.16 -0.10 & 103.8 & no \\
15 & 345.2216308 & 30.1636408 & 0.7 & 0.10  & 0.4 & 0.07 & 1.0 & 0.19 & 0.7 & 0.12 +0.11 -0.06 & 110.2 & no \\
16 & 345.2878914 & 30.1178499 & 1.0 & 0.15  & 5.2 & 0.98 & 3.8 & 0.71 & 3.7 & 0.65 +0.20 -0.16 & 153.8 & yes \\

    \noalign{\vskip 1mm}    
 \hline
    \noalign{\vskip 1mm} 
 \end{tabular}
 }
\footnotesize
\\
\flushleft
$^{a}$ Net count rate ($\times 10^{-4}\ \cts$). \\
$^{b}$ Broad band luminosity ($\times 10^{38}\ \ergs$). Assumes a power law with a photon index, $\Gamma=1.6$ and a distance to NGC~7457 of 12.1~Mpc. \\
$^{c}$ Projected distance from the centre of the galaxy (arcsec). \\
$^{d}$ Indicates whether the source has an optical counterpart in the \hst F475W and/or F850LP images. \\
* Source 01 is the nuclear source, its \lx measurements are based on fitting its spectra (rather than assuming $\Gamma=1.6$, see Section \ref{sec:agn})
\end{table*}

\chandra has observed NGC~7457 with the Advanced CCD Imaging Spectrometer (ACIS) on three epochs under proposal numbers 11900514 \citep[PI Gultekin, see][]{Gultekin12} and 16620677 (PI Peacock). We utilize these three observations: Obs ID 11786, 29~ks obtained on 11 September 2009; Obs ID 17007, 45~ks obtained on 11 September 2015; and Obs ID 18440, 51~ks obtained on 17 September 2015. 

We reduce and analyze these data using the Chandra Interactive Analysis of Observations ({\sc CIAO}) software \citep{Fruscione06}. All three observations are reprocessed using the {\sc chandra\_repro} script. We use {\sc dmextract} to extract the counts in background regions of these `evt' file. This identifies no flaring events in the three observations. All three observations have similar aimpoints (within $3\arcsec$). We utilize only data from the back illuminated S3 chip, which encloses the entire $r_{ext}$\footnote{The $r_{ext}$ ellipse is the ``total" aperture defined in the 2MASS large galaxy atlas, see \citet{Jarrett03} for details. It is a similar scale to the optical D25 ellipse. For NGC~7457 this ellipse has semi-major axis $= 155\arcsec\sim9.1$~kpc and ellipticity = 0.55} ellipse of the galaxy. 

Before analyzing these data, we align the different observations to a common reference system. To do this, we use {\sc fluximage} to produce exposure corrected images in the energy range 0.5-7.0 keV and {\sc mkpsfmap} to produce psf maps with an effective energy of 2.3 keV and an enclosed counts fraction of 0.9. Sources are identified in these broad band images using {\sc wavdetect} with $scales=1, 2, 4, 6, 8, 12, 16, 24, 32$ and detection threshold, $sigthresh=10^{-6}$. We select only well centred sources by requiring the sources have $counts>10$ and $psf\_size<7$~pixels. The three observations are aligned to the longest exposure observation (Obs~ID 18440) using {\sc wcs\_match} to calculate the transformations. The derived shifts ($\Delta ra,\Delta dec$) in are ($0.09\arcsec, 0.41\arcsec$) for Obs~ID 11786 and ($0.44\arcsec, 0.23\arcsec$) for Obs~ID 17707. These are applied to the event files (evt2) and aspect solutions (asol) using {\sc wcs\_update}.

 \begin{figure*}
 \centering
 \includegraphics[width=180mm,angle=0]{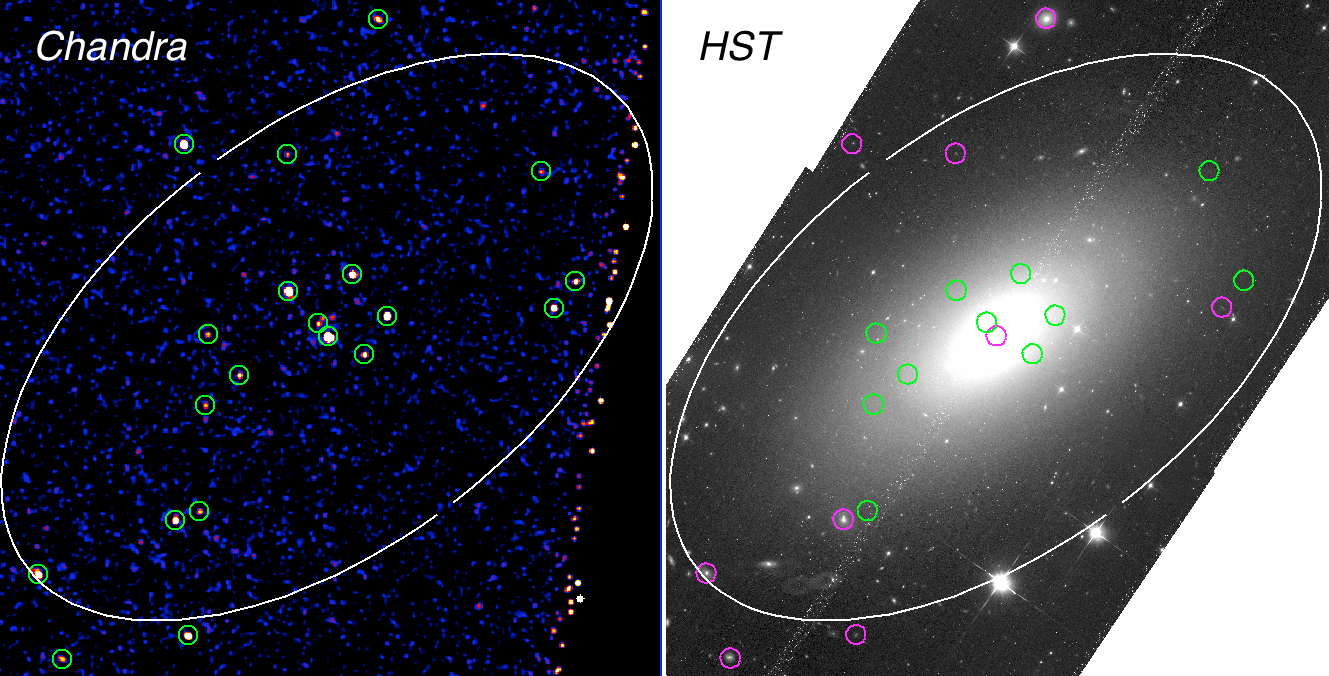} 
 \caption{LEFT: Merged \chandra broad band ACIS-S image of NGC~7457. The green circles show all of the sources detected by wavdetect. RIGHT: 2x1 \hst/ACS F850LP optical mosaic of NGC~7457. The magenta and green circles indicate the locations of X-ray sources with and without optical counterparts, respectively. In both frames the white line is the $r_{ext}$ ellipse of the galaxy, to which we restrict our analysis. This ellipse has a semi-major axis = 155$\arcsec \sim 9.1$~kpc and an ellipticity=0.55. We note that the four X-ray sources outside of this ellipse have optical counterparts and are likely background AGN. }
 \label{fig:images} 
\end{figure*}

Exposure corrected images in the \chandra source catalog's broad band ($0.5-7.0$~keV) are produced for the individual observations and the combined aligned data using {\sc merge\_obs}. These images are created with $binsize=1$, and hence maintain the native pixel size of 0.492$\arcsec$. PSF maps are created for the individual observations using {\sc mkpsfmap} with an effective energy of 2.3 keV and an enclosed counts fraction of 0.9. Such a map can not be constructed directly from the combined image. Instead, we construct a combined PSF map as an exposure weighted sum of the individual observation PSF maps. Sources are detected in the individual and combined observations using {\sc wavdetect} with $scales=1, 2, 4, 6, 8, 12, 16, 24, 32$ and detection threshold, $sigthresh=10^{-6}$. The combined 124.3~ks dataset reaches on an (on-axis) detection limit\footnote{Based on detecting eight counts and estimated using http://cxc.harvard.edu/toolkit/pimms.jsp} of $\lx \sim 1 \times 10^{37} \ergs$. We identify 16 sources within the $r_{ext}$ ellipse of NGC~7457. No transients are identified in the individual images that are not recovered in the combined image. 

For all of the  {\sc wavdetect} locations, source fluxes are extracted from circular regions enclosing 90$\%$ of the energy at 2.3~keV. Background regions are defined as an annulus $2-10$ times the source radius. These regions are created using {\sc psfsize\_srcs} and {\sc roi}, which also excludes other sources located in the background regions. We use these regions and the task {\sc srcflux} to determine the net counts and fluxes at the locations of each source in the combined data and in each of the three observations. Within this task, we use the arfcorr method to model the PSF and calculate the enclosed energy fraction for both the source and background regions. {\sc srcflux} also uses {\sc modelflux} to determine unabsorbed model fluxes for each source. We correct for Galactic $N_{\rm H}$\footnote{Obtained from http://cxc.harvard.edu/toolkit/colden.jsp for the centre of the NGC~7457 (23 00 59.9,  +30 08 42.0). }, assuming a single value of $5.5\times10^{20} {\rm cm}^{-2}$. Most sources have very few counts, we therefore model the flux from all sources as a single power law with a photon index of 1.6. Finally, we convert fluxes to luminosities assuming a distance to NGC~7457 of 12.1~Mpc \citep{Tully13}. 

To measure faint sources and improve the signal to noise, we also require source fluxes measured from the combined dataset. However, we cannot simply obtain these by running {\sc srcflux} on the combined event file. Instead, we choose to produce combined fluxes by combining the results from our individual analysis. For each source, we use {\sc spectra\_combine} to combine the pulse invariant (`pi') and response (`nopsf.arf') files produced for each source and calculate the exposure weighted net count rate from the individual `NET\_RATE\_APER' values determined by {\sc srcflux}. We then use {\sc modelflux} with these combined arf files, rmf files and weighted net count rates to determine the unabsorbed model fluxes, assuming the same models and $N_{\rm H}$ discussed above. Uncertainties on the combined flux are calculated by propagating the errors on the individual observations. 

The detected sources, their count rates, and luminosities are listed in Table \ref{tab:data}.

\section{\hst observations}
\label{sec:optical}

The X-ray sources detected within the $r_{ext}$ ellipse of NGC~7457 are likely to be a combination of LMXBs in the field of the galaxy, LMXBs within its globular clusters, and contamination from background AGN/ foreground stars. We can distinguish between these sources by identifying optical counterparts to globular cluster sources and background AGN. To do this we utilize \hst Advanced Camera for Surveys (ACS) observations of the galaxy. A 2x1 mosaic, which covers the entire $r_{ext}$ ellipse of NGC~7457, was obtained on 2014-12-11 (Program ID 13942; PI Peacock). The two fields were observed through the F475W and F850LP filters (similar to the ground based $g$- and $z$-bands) with total exposure times of 2204s and 2100s, respectively. These \hst images and the \chandra ACIS-S3 broad band image are shown in Figure \ref{fig:images}. Detected X-ray sources are highlighted. It can be seen that some X-ray sources appear to be aligned with optical sources (magenta circles in Figure \ref{fig:images}). 

We utilize the pipeline reduced `flc' images, downloaded from the MAST archive\footnote{archive.stsci.edu}. We align these images to a common reference frame using the {\sc drizzlepac} task {\sc tweakreg} and combine the four observations through each filter to produce a stacked mosaiced image using {\sc astrodrizzle}. The two mosaiced images are then aligned to the \chandra source catalog by matching seven sources (including five outside of $r_{ext}$) to optical counterparts using the {\sc iraf} task {\sc msctpeak}. The galaxy light is subtracted from these images using the {\sc iraf} task {\sc rmedian}, which applies a ring median filter to the images with inner and outer radii of 30 and 31 pixels, respectively. Photometry is performed on these background subtracted images through an aperture with a radius of $0.25\arcsec$ using {\sc sextractor} \citep{Bertin96}. Aperture corrections from 0.25$\arcsec$ to 0.5$\arcsec$ are calculated by comparing photometry through such apertures for bright sources. An additional correction from 0.5$\arcsec$ to $\infty$ is applied based on the corrections presented by \citet{Bohlin11}. The resulting F475W and F850LP catalogs are combined using {\sc topcat} \citep{Taylor06}.

\subsection{NGC~7457's globular clusters} 
\label{sec:gcs}

 \begin{figure}
 \centering
 \includegraphics[width=88mm,angle=270]{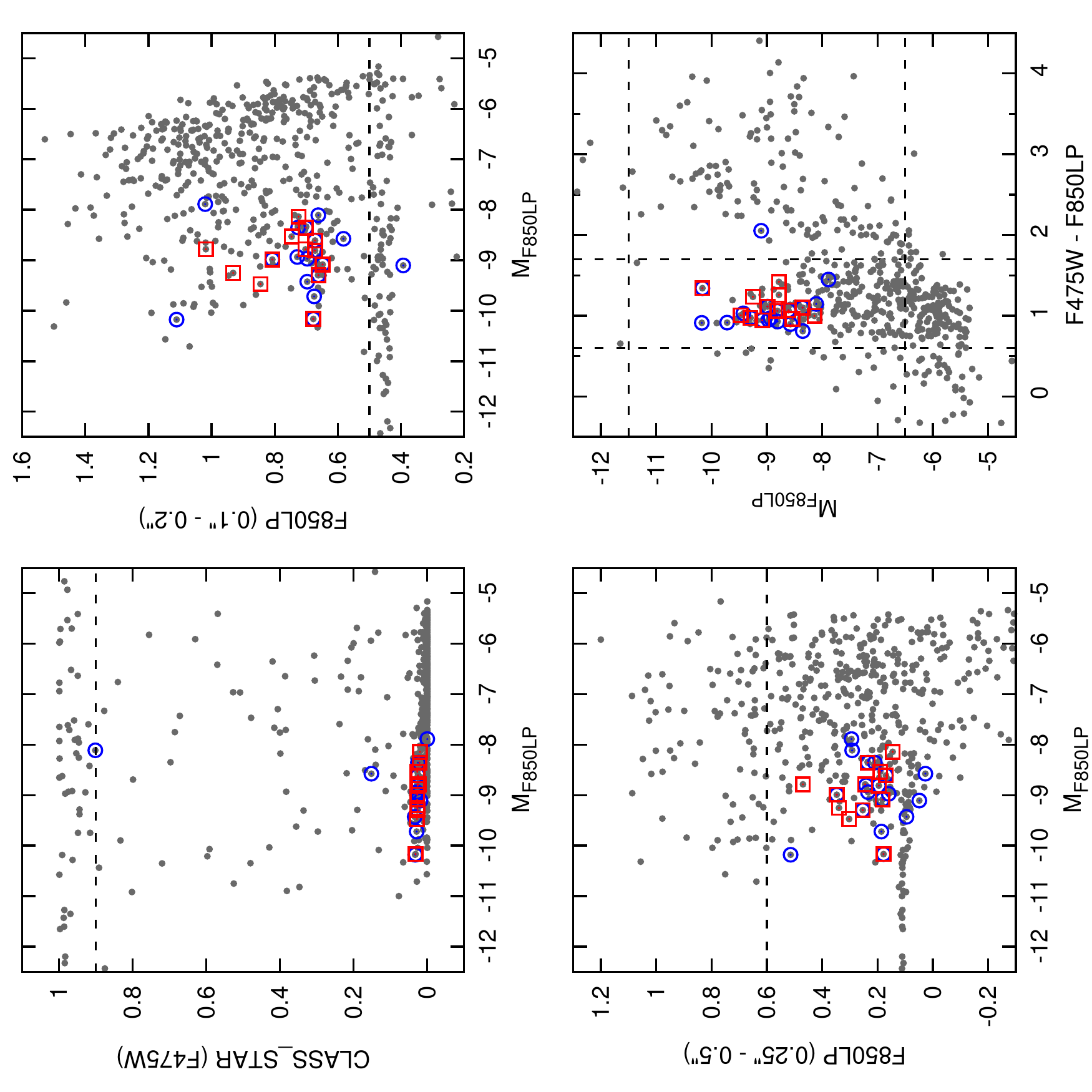} 
 \caption{Properties of all sources detected in these F475W and F850LP images (grey points). The dashed black lines indicate our globular cluster selection criteria. The properties of sources previously identified as globular clusters from spectroscopic observations are highlighted with red squares and blue circles. See Section \ref{sec:gcs} for details. }
 \label{fig:gc_selection} 
\end{figure}

NGC~7457 is known to host a small globular cluster system, estimated at $210\pm30$ clusters \citep{Hargis11}. A sample of these clusters have been spectroscopically confirmed and studied \citep{Chomiuk08, Pota13}. A comprehensive analysis of this globular cluster system is beyond the scope of this study. However, we do wish to identify globular clusters in these \hst data in order to classify globular cluster LMXBs. In this section, we therefore present our cluster selection criteria and give a brief discussion of their properties. 

To identify globular cluster candidates from this \hst photometry, we require that they: (1) have $-11.5<M_{\rm F850LP}<-6.5$, which corresponds to stellar masses of $4\times10^{4} \lesssim M_{\star} \lesssim4\times10^{6}\Msun$ (assuming $M_{z\odot}=4.5$ and $M_{\star}/L_z=1.7 M_{\odot}/L_{z\odot}$, based on the models of \citealt{Maraston05} for a 12~Gyr population with $[Z/H]=-1.35$ and a Kroupa IMF); (2) have $0.6<{\rm F475W-F850LP}<1.7$, as observed in other globular cluster systems \citep[e.g.][]{Peng06, Strader06}; (3) are extended, with {\sc class\_star}~$<0.9$ and ${\rm F850LP(0.1\arcsec)-F850LP(0.2\arcsec)}>0.5$; (4) they are not too extended to be a cluster at this distance, with ${\rm F850LP(0.25\arcsec)-F850LP(0.5\arcsec)}<0.6$. Figure \ref{fig:gc_selection} shows these selection criteria and all sources detected in the radial region $10\arcsec<r<r_{ext}$ (grey points). We have highlighted previously proposed globular cluster candidates that have been spectroscopically confirmed by \citet[][red squares]{Chomiuk08} and \citet[][blue circles]{Pota13}. All of these spectroscopic clusters are detected, although one is located in the noisier chip gap region of our F850LP photometry and its properties are significantly effected by a cosmic ray. The other spectroscopic clusters are all classified as globular clusters by our analysis. 

In Figure \ref{fig:gz_z}, we present the observed CMD of all sources (grey points) and our globular cluster candidates (black points). We identify 65 globular cluster candidates with $M_{F850LP}<-8.0$. We estimate the total number of clusters in our radial region by binning these 65 clusters into 0.5 mag bins and fitting the luminosity function to a lognormal distribution. Because of the small number of clusters, we fix the peak of the distribution at $M_{F850LP}=-8.46$ and the dispersion at 0.87 \citep[consistent with the GCLF study of][]{Villegas10}. We calculate a total of 121 globular clusters in $10\arcsec<r<r_{ext}$. This corresponds to a mass scaled specific frequency of globular clusters \citep[e.g.][]{Zepf93}, $T=N_{GC}/(M_{\star}/10^{9}\Msun)=9.5$. We take the stellar mass covered by our study to be $M_{\star}=12.8\times10^{9}\Msun$, derived from the $K$-band light covered \citep[calculated as $90\%$ of the total \lk based on the 2MASS LGA images,][]{Jarrett03} by assuming $M_{\star}/L_{K}=0.85$ \citep{Fall13}. We note that this is significantly larger than the value derived by \citet{Hargis11}. This is primarily due to their use of a large $M_{\star}/L_{V}$ which predicts a significantly higher mass for the galaxy. We use the $K$-band derived mass since the variation in $M_{\star}/L_{K}$ is known to be relatively small \citep[e.g.][]{Bell01}. This $T$ is also directly comparable to that previously determined for more massive galaxies by \citet{Peacock16}, who find $5<T<30$, suggesting that the small globular cluster population is consistent with the galaxy's relatively low mass. 

The globular cluster systems of most early-type galaxies show clear colour bimodality \citep[e.g.][]{Brodie06}. However, previous studies have found no clear evidence for this in NGC~7457's clusters. \citet{Hargis11} presented ground based ${\rm B-R}$ colors of these clusters and find that they are consistent with the (bimodal) colour distribution of the Milky Way's globular clusters. However, they found no significant evidence for colour bimodality \citep[see also][]{Pota13}. \citet{Chomiuk08} also found no significant evidence for bimodality in the ${\rm V-I}$ colours of these clusters, based on the smaller area of a single WFPC2 field. We test for bimodality in the $g-z$ colours of our 65 globular cluster candidates with $M_{F850LP}<-8.0$. To do this, we use the KMM test \citep{Ashman94}, implemented using the GMM code of \citet{Muratov10}. We run this test for two Gaussians with equal variance and obtain a best fit variance, $\sigma_{12}=0.10$ and peaks in $g-z$ at $\mu_{1}=1.00$ and $\mu_{2}=1.27$. The bimodal distribution is marginally favoured, with the unimodal distribution rejected with $P$-value~=~0.05. The peaks are also significantly separated with $D=(\mu_{1}-\mu_{2})/\sigma_{12}=2.8$, where $D>2$ is required to cleanly separate the two peaks \citep{Ashman94}. The evidence for bimodality in our data may be due to our increased coverage and hence larger sample size than \citet{Chomiuk08}, while the higher spatial resolution of these \hst data may result in lower contamination from non-cluster sources than the study of \citet{Hargis11}. However, we caution that the bimodal case is only somewhat favoured and the number of clusters (65) is only slightly above the number required for the KMM test to be reliable \citep[][]{Ashman94}.

\subsection{Optical counterparts to X-ray sources} 

 \begin{figure}
 \centering
 \includegraphics[width=85mm,angle=0]{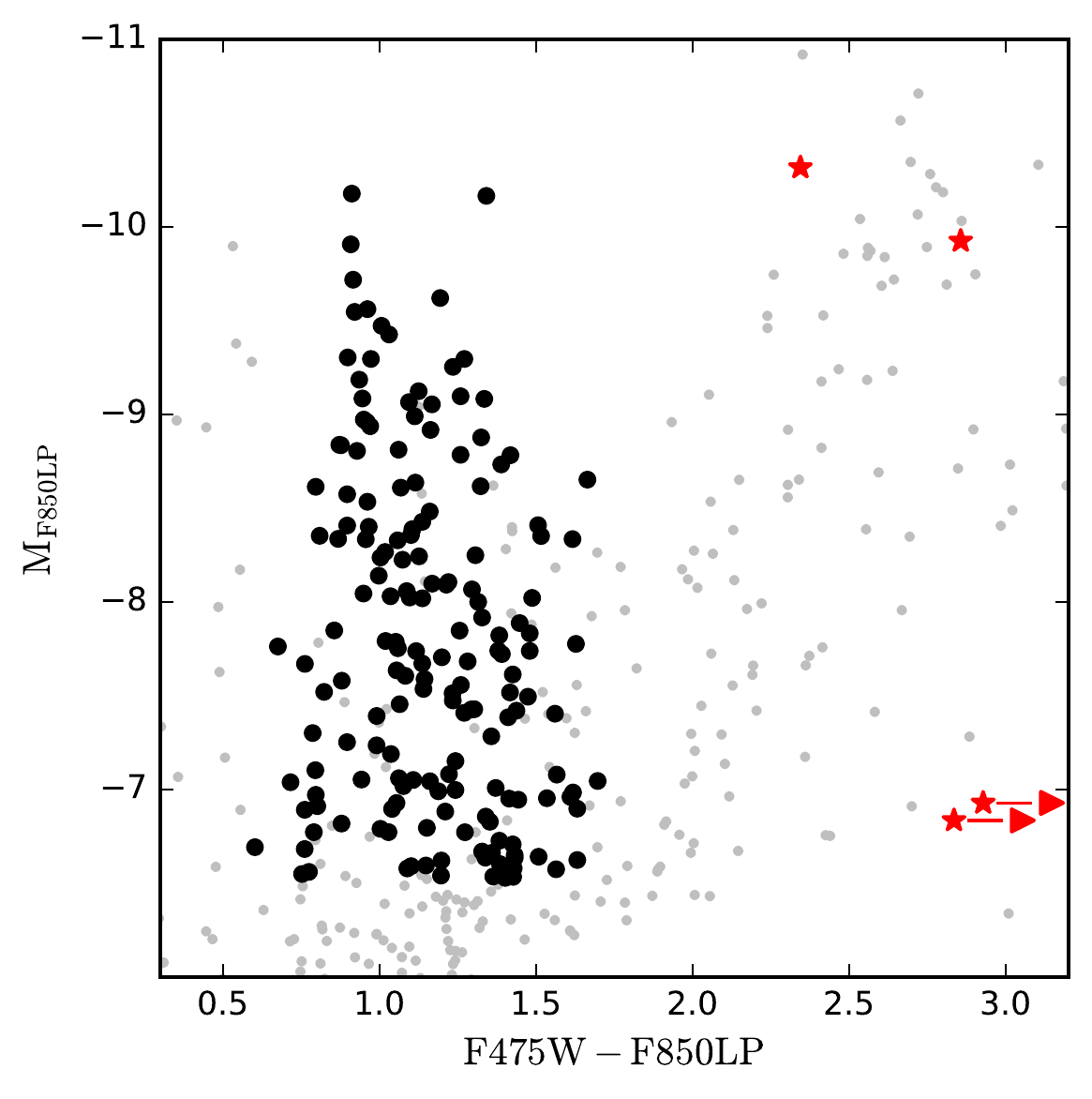} 
 \caption{Colour magnitude diagram of optical sources identified within the $r_{ext}$ ellipse of NGC~7457. Globular cluster candidates are highlighted as solid black circles. The red stars show sources with X-ray emission. The red colours of these X-ray sources are consistent with them being background galaxies. }
 \label{fig:gz_z} 
\end{figure}

 \begin{figure}
 \centering
 \includegraphics[width=85mm,angle=0]{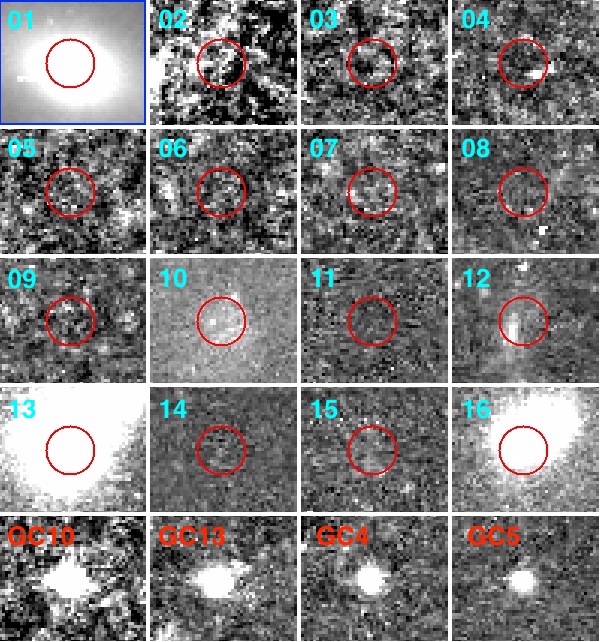} 
 \caption{F850LP images at the location of each X-ray source (from top left to bottom right, sources $01-16$ from Table \ref{tab:data}). Red circles have radii~$ = 0.4\arcsec$ and indicate the location of each X-ray source. For comparison, the bottom row shows four clusters from the catalog of \citet{Chomiuk08}. From left to right, these clusters have $M_{F850LP}=-10.18, -9.72, -8.57, -8.11$. All images are background subtracted and have the same zscale (except the central source, whose scale is adjusted to show the peak emission). }
 \label{fig:thumbnails} 
\end{figure}

Our final optical catalog is matched to the X-ray source catalog using {\sc topcat}, based on the aligned WCS and a matching radius of 0.4$\arcsec$. This identifies four sources within $r_{ext}$ that have optical counterparts. In Figure \ref{fig:thumbnails}, we show F850LP cutouts for each X-ray source location. Source 01 is aligned with the center of the galaxy. This nuclear source is discussed in Section \ref{sec:agn}. Source 02 is also close to the centre of the galaxy and is removed from our analysis because optical counterpart matching is less reliable in this inner region. It can be seen that four other sources have optical counterparts. Two have bright and extended counterparts (13 and 16). The other two have much fainter, but also extended counterparts (10 and 12). For comparison we also show F850LP cutouts of four spectroscopically confirmed globular clusters, which span the range of magnitudes studied by \citet{Chomiuk08}. 

Figure \ref{fig:gz_z} shows the colour magnitude diagram for all of our sources (grey points), our globular cluster candidates (black points) and these four sources which have X-ray counterparts (red stars). Two of the X-ray counterparts are faint and red and therefore only detected in F850LP. We plot upper limits on the colours of these sources. All four of these sources are too red to be globular clusters in NGC~7457 and they are likely to be AGN in background galaxies. These sources are flagged in Table \ref{tab:data} and are excluded from our analysis.

\section{Field LMXB population} 
\label{sec:lmxb}

 \begin{figure}
 \centering
 \includegraphics[width=88mm,angle=0]{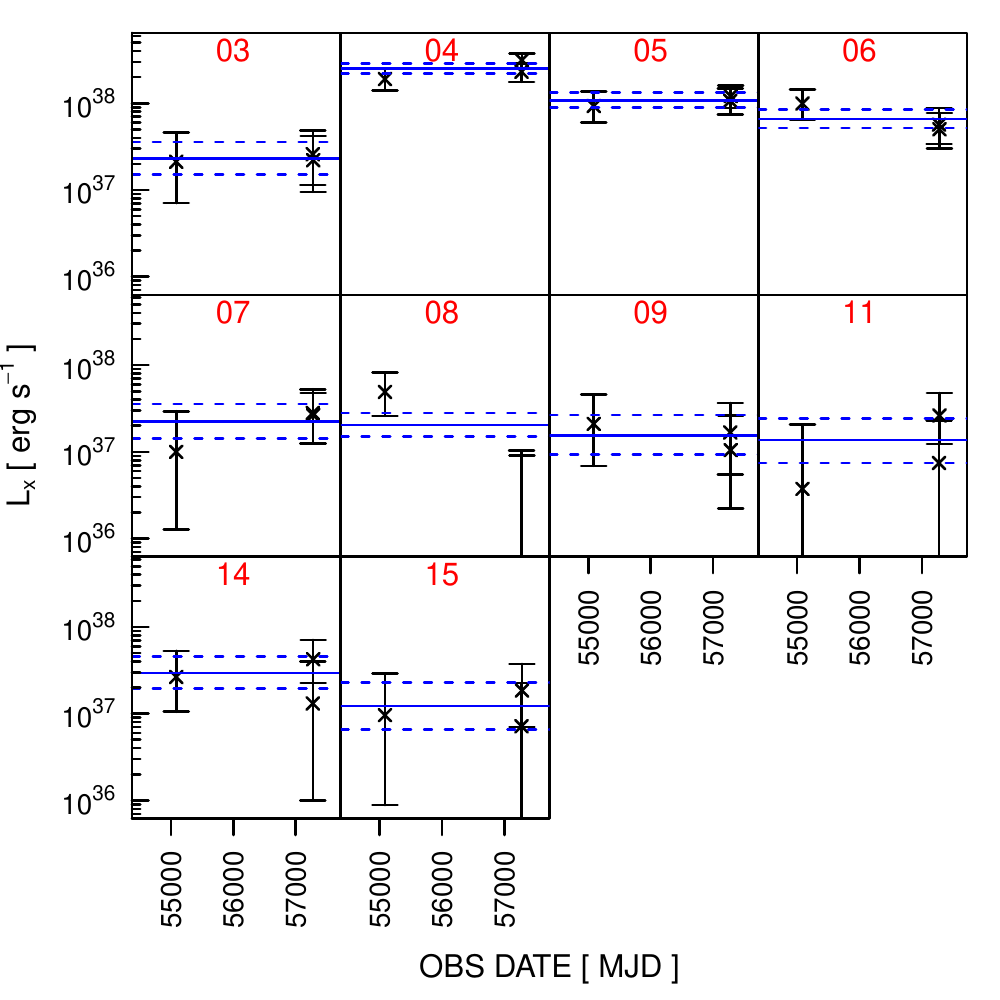} 
 \caption{Luminosities of all field LMXBs detected within $r_{ext}$. Black points show \lx at the three epochs available. The solid (and dashed) blue lines show the average luminosity (and errors) for the combined data. The red source IDs correspond to those in Table \ref{tab:data}. Source 08 is only detected in the first epoch, upper limits are shown for the later observations. }
 \label{fig:light_curves} 
\end{figure} 

Ten X-ray sources have no optical counterparts and are classified as field LMXBs in NGC~7457. In Figure \ref{fig:light_curves}, we plot the luminosity of these LMXBs as a function of observation date. The blue lines show the combined luminosity and uncertainty. Only one transient is observed. Source 08 is detected in 2009, but not in the later 2015 observations. 

\begin{figure}
 \centering
 \includegraphics[width=88mm,angle=0]{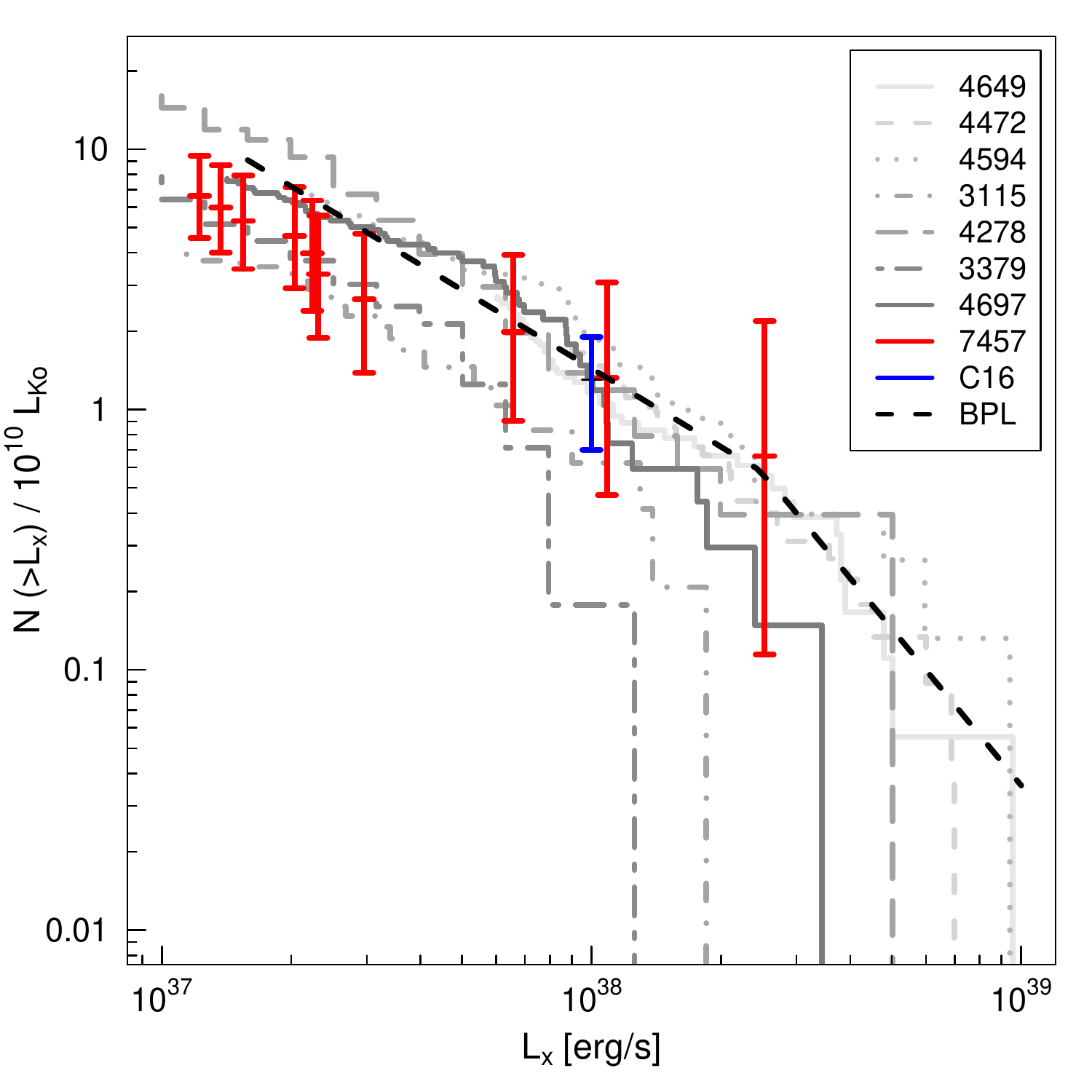} 
 \caption{The cumulative X-ray luminosity function of field LMXBs in NGC~7457 (red points) and the average of other similar mass galaxies \citep[from][C16; blue point]{Coulter16}. They grey lines show the XLF of other, more massive, early-type galaxies. Darker lines correspond to lower velocity dispersion galaxies. The black-dotted line is the broken power law (BPL) representation of the XLFs, scaled to fit all of the galaxy data. }
 \label{fig:xlf} 
\end{figure}

In Figure \ref{fig:xlf}, we show the X-ray luminosity function (XLF) of the field LMXBs identified in NGC~7457 (red points). For bright LMXBs (with $\lx > 10^{38} \ergs$), we also include the average number detected in this, and five other, low mass early-type galaxies (with $78<\sigma<110$~\kms; blue point). This point is taken from \citet{Coulter16} and is based on the detection of 12 LMXBs in shallower \chandra observations of NGC~4339, NGC~4387, NGC~4458, NGC~4550, NGC~4551 and NGC~7457. The bright LMXB populations of these similar mass early type galaxies are in good agreement with NGC~7457's XLF. 

We compare NGC~7457's field LMXB XLF to those published in other, more massive early-type galaxies (with $180<\sigma<305~\kms$; grey lines in Figure \ref{fig:xlf}). These data are taken from \citet[][]{Peacock14} for NGC~1399, NGC~3379, NGC~4278, NGC~4472, NGC~4594, NGC~4649, NGC~4697 and \citet[][]{Lehmer14} for NGC~3115. Globular cluster LMXBs and background AGN have been removed from these data in a similar way to our analysis of NGC~7457; by matching to optical counterparts in \hst images. We scale the XLFs by the amount of $K$-band light covered, based on the $K_{ext}$ magnitude from the 2MASS LGA \citep{Jarrett03} and the fraction of $K$-band light covered by these X-ray catalogs (taken from \citealt{Peacock14} and \citealt{Lehmer14}). For NGC~7457, the region studied ($10\arcsec<r<r_{ext}$) covers $90\%$ of the $K$-band light. 

The black-dashed line in Figure \ref{fig:xlf} shows a broken power law with an exponent $\alpha_{1}=-1.0$ below a break at $\lx=2.5\times10^{38}\ergs$ and an exponent of $\alpha_{2}=-2.0$ above this break. This has previously been shown to provide a good representation of the field LMXB XLF \citep[e.g.][]{Kim09, Kim10, Zhang11, Lehmer14, Peacock16}. Scaling to fit all of these galaxies, we find that the number of LMXBs with $\lx>2\times10^{37}\ergs$  per $10^{10} \LKsun$, $n_{x} \sim 7$. 

NGC~7457's field LMXB population is found to be similar to that of these other galaxies with $n_{x}=4.6^{+2.5}_{-1.7}$. The similarity of the number of field LMXBs per \lk at all luminosities suggests a similar formation efficiency of LMXBs across these galaxies. Of particular note is that no systematic difference is observed due to the mass of these galaxies, which span a wide range of velocity dispersions: $78<\sigma<305~\kms$. This suggests that factors such as primordial binary fraction and initial mass function (IMF) do not vary significantly with galaxy mass, or that they vary in a way that does not influence LMXB formation. 

 \begin{figure*}
 \centering
 \includegraphics[width=130mm,angle=0]{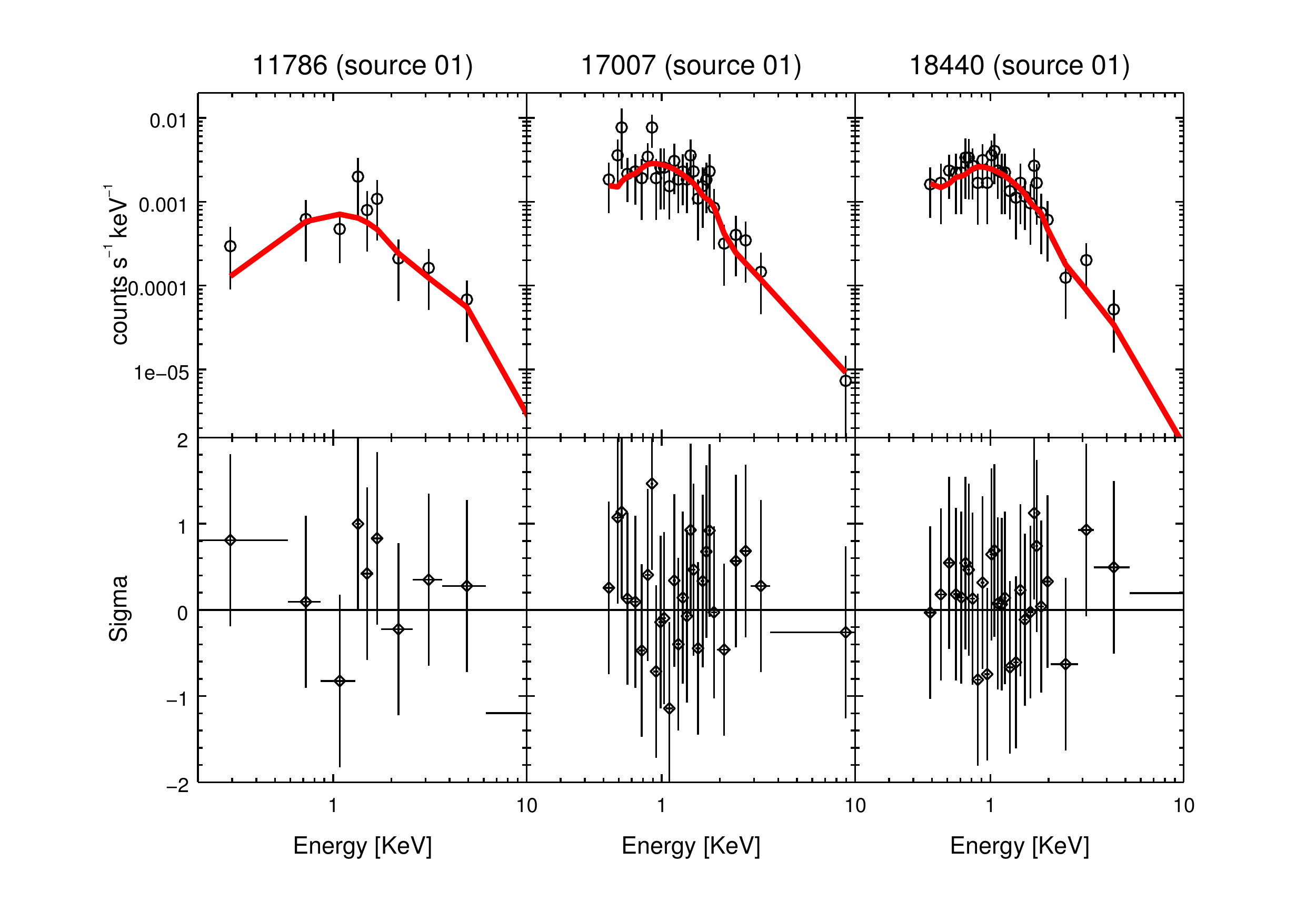} 
 \caption{(top) Spectra of the nuclear X-ray source in our three datasets (from left to right: 11786; 17007; 18840). Red lines show the best fitting absorbed powerlaw models. (bottom) Residuals for these best fit models. }
 \label{fig:nuclear} 
\end{figure*}

\section{Globular cluster LMXB population} 
\label{sec:gc_lmxb}

The formation of LMXBs is known to be much more efficient in globular clusters than in the fields of galaxies. Indeed observations of other early-type galaxies have shown that large fractions of the LMXB populations are located in their globular clusters \citep[e.g.][]{Angelini01, Kundu02}. Therefore, it may seem surprising that we detect no LMXBs in NGC~7457's globular clusters. 

To investigate the significance of the non-detection of LMXBs in NGC~7457's clusters, we empirically predict the number of globular cluster LMXBs expected based on observations of NGC~3379, NGC~4278 and NGC~4697. Data for these galaxies were presented in \citet{Peacock14}. Their LMXB populations have been observed to similar detection limits to NGC~7457 and they have similar \hst F475W and F850LP data that have been used to study their globular cluster LMXBs. We restrict the sample to clusters with $M_{F850LP}$~$<-7.5$ and estimate the stellar mass of the clusters assuming a $M_{z\odot}=4.50$ and a mass-to-light ratio, $M_{\star}/L_z=1.7 M_{\odot}/L_{z\odot}$ (based on the models of \citealt{Maraston05} for a 12~Gyr population with $[Z/H]=-1.35$ and a Kroupa IMF). There is known to be an enhancement in the presence of LMXBs with metallicity \citep[e.g.][]{Kundu07, Sivakoff07, Kim13}, we therefore also split the samples into red (with F475W~--~F850LP~$>$~1.2) and blue globular clusters. 

We consider globular cluster LMXBs with $\lx>2\times10^{37}\ergs$ (all of which should be detected in these data). We find that red (and blue) globular clusters host 0.12 (0.02), 0.11 (0.04), 0.10 (0.05) LMXBs per $10^{6}\Msun$ in NGC~3379, NGC~4278 and NGC~4697, respectively. Therefore, on average, 0.109 and 0.033 LMXBs are expected per $10^{6}\Msun$ in red and blue globular clusters, respectively. To compare with the fields of these galaxies, we convert this to the number per $K$-band luminosity assuming $M_{\star}/L_{K}=1.0 M_{\odot}/L_{K\odot}$ \citep[using the models of][discussed above]{Maraston05}. We find that 1090 (and 330) LMXBs are observed per $10^{10} \LKsun$ in the metal rich (and poor) globular clusters. Comparing with the number of LMXBs observed in the fields of these galaxies (4-12 per $10^{10} \LKsun$, see Section \ref{sec:lmxb}) highlights the enhanced formation efficiency of bright LMXBs in globular clusters. 

This estimate of the number of LMXBs is based on the stellar mass of the globular clusters. In fact, we know that the dominant parameter that drives the formation of LMXBs in globular clusters is the stellar interaction rate \citep[e.g.][]{Verbunt87, Pooley03, Peacock09}. Unfortunately, measuring this parameter for clusters beyond the local group is extremely challenging as it involves resolving the clusters to their core radii. However, the mass of a cluster is correlated with interaction rate \citep[e.g.][]{Davies04, Peacock10b}. Therefore, if the statistical sample is large enough, the number of LMXBs should scale with stellar mass. 

As discussed in Section \ref{sec:gcs}, the number of globular clusters in NGC~7457 is quite small, consistent with the relatively low mass of the galaxy. We detect 99 globular clusters with $M_{F850LP}<-7.5$. Assuming $M_{\star}/L_z=1.7 M_{\odot}/L_{z\odot}$ (see above), these red (and blue) globular clusters have a total stellar mass of $9.0\times10^{6}\Msun$ (and $21.4\times10^{6}\Msun$). From the average numbers of LMXBs per stellar mass, calculated above, this predicts that NGC~7457's clusters should host 1.7 LMXBs with $\lx>2\times10^{37}\ergs$ or $\sim3$ above our detection limit of $\lx>1\times10^{37}\ergs$ \citep[based on the observed XLF of globular clusters;][]{Kim10, Peacock16}. Therefore, while a few LMXBs would be expected on average, the non-detection in these clusters is not that surprising given the small numbers involved and the uncertainty in the interaction rates of these clusters.

\section{Nuclear X-ray emission} 
\label{sec:agn} 

The brightest X-ray source in our sample (source 01 in Table \ref{tab:data}) is well aligned with the optical centre of the galaxy. In Figure \ref{fig:nuclear} we plot the spectra of this source for the three separate observations. Each spectrum is independently fit to absorbed power law models using {\sc sherpa}. Unabsorbed model fluxes are then calculated using {\sc modelflux}, as in Section \ref{sec:xray} (but using the derived photon indices, $\Gamma$). Observations 17007 and 18440 were taken at similar times and show similar properties with $\Gamma=2.9\pm0.3$ and $3.0\pm0.3$ and $\lx=(6.8$ and $5.7)\times10^{38}\ergs$, respectively. Observation 11786 was taken six years earlier and shows harder and fainter emission, with $\Gamma=1.5\pm0.5$ and $\lx=2.8\times10^{38}\ergs$. 

The central SMBH in NGC~7457 is one of only a small number with direct mass determinations. From modeling of the stellar kinematics, the mass of this black hole was calculated to be $4.1^{+1.2}_{-1.7} \times10^{6}\Msun$ \citep{Gebhardt03, Gultekin09a}. From this, we estimate the Eddington luminosity,  $L_{Edd}\sim 1.26\times 10^{38} M/\Msun = 5.1\times 10^{44} \ergs$. 

Using these data, we calculate that the ratio of X-ray luminosity ($0.5-7.0$~KeV band) to Eddington luminosity ($\lx/L_{Edd}$) increases from $0.5\times10^{-6}$ to $1.3\times10^{-6}$. We note that \citet{Gultekin12} previously presented $\lx/L_{Edd}$ based on one of these observations, 11786. Their measurement was a factor of two lower, although our studies do not apply a bolometric correction, so this could be due to their use of a different energy range ($2-10$~KeV). Interestingly, NGC~7457 has the lowest mass SMBH of the 12 galaxies in the sample of \citet{Gultekin12} and, for the increased flux of the source in the 2015 observations, it has highest $\lx/L_{Edd}$ ratio. This is consistent with some other observations and the theory of ``down-sizing", where lower mass black holes may accrete closer to their Eddington luminosity than higher mass black holes \citep[e.g.][]{Gallo10}. 

\section{Conclusions}
\label{sec:conclusions}

We investigate the X-ray populations of NGC~7457 based on deep \chandra observations. We identify 10 X-ray point sources in the $r_{ext}$ ellipse of the galaxy which have no optical counterparts and are likely LMXBs in the field of NGC~7457. We show that this small field LMXB population is consistent with the populations observed in other early-type galaxies, given the relatively low stellar mass of NGC~7457. We also show that its XLF is similar to that of these other (more massive) early-type galaxies, both in shape and scale. Our results suggests a similar formation efficiency of LMXBs across a broad range of host galaxy mass and environment. 

From \hst observations of the galaxy we estimate that 121 globular clusters are present in the region studied $10\arcsec<r<r_{ext}$. We  find that the cluster system is somewhat bimodal, similar to that of other early type galaxies. Previous work based on ground based data and \hst WFPC2 data for a smaller region did not find evidence for such bimodality, possibly due to smaller samples or increased contamination. 

None of NGC~7457's globular clusters are found to host LMXBs with $\lx>10^{37} \ergs$. Considering other galaxies with similarly deep X-ray observations, we calculate the number of globular cluster LMXBs expected per stellar mass. We find that on average 0.11 and 0.03 LMXBs are observed per $10^{6}\Msun$ in the red and blue globular clusters of these galaxies, respectively. This predicts around 3 LMXBs in NGC~7457's globular cluster system. Given the small numbers and uncertainties in this prediction, this is consistent with our non-detection. 

Nuclear X-ray emission is detected from the relatively low mass central SMBH. We calculate that the ratio of X-ray to Eddington luminosity of this source increases from $0.5\times10^{-6}$ to $1.3\times10^{-6}$.  This is the highest $\lx/L_{Edd}$ ratio among the 12 galaxies with ``secure" SMBH masses studied by \citet{Gultekin12}. This is consistent with the idea that lower mass black holes may accrete at higher Eddington fractions than higher mass black holes.

\section*{Acknowledgements}

We thank the referee of this paper, Paolo Bonfini, for providing detailed and informed feedback on the original manuscript. We are very grateful to the CXC helpdesk for providing detailed and helpful advice on the analysis of these merged ACIS data. This research has made use of software provided by the Chandra X-ray Center (CXC) in the application packages {\sc CIAO}, {\sc ChIPS}, and {\sc Sherpa}. This research has made use of NASA's Astrophysics Data System.

The scientific results reported in this article are based on observations made by the Chandra X-ray Observatory. Support for this work was provided by NASA through Chandra Award Numbers GO5-16084A (MBP) and GO5-16084B (AK) issued by the Chandra X-ray Observatory Center, which is operated by SAO for and on behalf of the National Aeronautics Space Administration under contract NAS8-03060.

Based on observations made with the NASA/ESA Hubble Space Telescope, obtained at the Space Telescope Science Institute, which is operated by the Association of Universities for Research in Astronomy, Inc., under NASA contract NAS 5-26555. These observations are associated with program HST-GO-13942.001-A. Support for Program number HST-GO-13942.001-A was provided by NASA through a grant from the Space Telescope Science Institute, which is operated by the Association of Universities for Research in Astronomy, Incorporated, under NASA contract NAS5-26555.

\bibliographystyle{mnras}

\label{lastpage}

\end{document}